\documentstyle[psfig,conf_iap,]{article}
\begin{document}
\heading{%
%
Could Dark Energy be Measured from Redshift Surveys ?\\
%
} 
\par\medskip\noindent
\author{%
Ofer Lahav$^{1}$
}
\address{%
Institute of Astronomy, University of Cambridge, \\
Madingley Road, Cambridge CB3 0HA \\
e-mail: lahav@ast.cam.ac.uk
}

\begin{abstract}
We review the ability of redshift surveys to provide constraints on
the Dark Energy content of the Universe.  The matter power spectrum
and dynamics at the present epoch are nearly `blind' to Dark Energy, but
combined with the CMB they can provide a constraint on the Equation of State
parameter $w$.  A representative result from the 2dF galaxy redshift
survey combined with the CMB is $w <-0.52$ (95\% CL; with a prior of
$w \ge -1$), consistent with
Einstein's Cosmological Constant model ($w=-1$).  More complicated forms of
Quintessence (e.g. epoch-dependent $w$ or $w<-1$) 
are not yet ruled out.  At higher
redshifts, the abundance of galaxies and clusters of galaxies, variants of the
Alcock-Paczynski curvature test and cross correlation of the CMB with
radio sources look potentially promising, but they suffer from
degeneracy with other parameters such as the matter density and galaxy
biasing.
\end{abstract}
\section{Introduction}

Recent cosmological observations suggest the existence of a Dark
Energy component in the Universe (e.g. Bahcall et al. 1999; Efstathiou
et al. 2002).  However, the data still allow room for a complicated
cosmic Equation of State $w = P/\rho$ which may vary with redshift, as
proposed by models of `Quintessence' (e.g. Sahni \& Starobinsky 2000;
Peebles \& Ratra 2002 for a review). 

We shall first clarify notation and convention in this field.
The case $w=-1$
corresponds to Einstein's Cosmological Constant.  
Then the only Dark Energy  parameter is the present-epoch 
Dark Energy density $\Omega_{Q0}$,
which commonly appears in the literature as
$\lambda \equiv \Lambda/(3 H_0^2)$ or $\Omega_\Lambda$.
If a general Equation of State is allowed then 
one has to solve for both $w(z)$ (parameterized e.g. as
$w(z) = w = const.$ or  $w(z) = w_0 + w_1 z$) 
as well as for $\Omega_{Q0}$ (see below).
In quoting  results from the literature we shall specify 
which model of Cosmological Constant or Quintessence is assumed
and what is the prior on the allowed range of $w$.
For example, $w \geq -1$  is 
expected for standard minimally coupled scaler fields (e.g. Sahni
\& Starobinsky 2000), although $w < -1$ is
allowed in some scenarios (e.g. Melchiorri et al. 2002).

Here we discuss what
can (or cannot) be learned from redshift surveys alone and with input
from other cosmic probes such as the Cosmic Microwave Background (CMB).  The
methods we consider include galaxy clustering and dynamics, the
Alcock-Paczynski curvature test, abundance of clusters and galaxies
with redshift, and cross correlation of the galaxy surveys with the
CMB to detect the integrated Sachs-Wolfe effect.

\section{Clustering and Dynamics}

It is worth emphasizing from the start that the
fundamental function that controls the growth of structure
is the epoch-dependent Hubble parameter: 

\begin{equation}
H^2(z)/H^2_0 = (1+z)^3 \Omega_{m0} +(1+z)^2 \Omega_{k0}
+ \Omega_{Q0} \exp [3 X(z)]  
\label{eq:H}
\end{equation}
where
\begin{equation}
X(z) = \int_0^z [1+w(z')] (1+z')^{-1} dz', 
\label{eq:X}
\end{equation}
and $\Omega_{m0}$ and $\Omega_{Q0}$ are the present density parameter
of matter and dark energy components and 
$\Omega_{k0} = 1- \Omega_{m0} - \Omega_{Q0}$ is the present curvature. 

\subsection{Growth of Perturbations and Biasing}

The growth of structure in the Universe  
is sensitive to the underlying cosmological model
due to the dependence on $H(z)$.
In linear theory the density contrast 
$\delta \equiv \frac{\delta \rho}{\rho}$ obeys
\begin{equation}
{\ddot{\delta} }+ 2 H {\dot{\delta}} = {3 \over 2} H_0^2 \Omega_{m0} (1+z)^3 \delta  
\label{eq:D}.
\end{equation}
Starobinsky (1998) evaluated 
$H(z)$ for a given $\delta(z)$. 


There are at least two problems in trying to deduce $w(z)$ from
surveys at different redshifts: (i) The function 
$w(z)$ appears under an integral
[eq. (2)], hence detailed information on the redshift dependence
of $w$ is `washed out'.  A similar problem
exists in trying to deduce $w(z)$ from the luminosity distance of
e.g. Supernova Ia (Maor et al. 2002);
(ii) Ideally we would like to map the growth of structure by observing
galaxy clustering at different redshifts.  However, the relation
between the galaxy density contrast $\delta_g(z)$ and the (growing
mode) mass density contrast $\delta(z)$ is non-trivial, and is usually
parameterized via the (linear) biasing parameter $b(z) =
\delta_g(z)/\delta(z)$.  Simple forms of redshift-dependent biasing
exist, e.g. by assuming that galaxies follow the cosmic flow as test
particles (Fry 1996):
\begin{equation}
b(z) = 1 + [b(0) - 1]/\delta(z) 
\label{eq:biasing}.
\end{equation}
Note that even if biasing was large at high redshifts it would tend to unity at the present epoch ($b \sim 1$ is supported by measurement 
of the biasing of bright
2dF galaxies on large scales, e.g. Lahav et al. 2002; Verde et al. 2002). 
In reality it is more complex as galaxies of different types cluster differently on
small scales, and hierarchical merging scenarios suggest a more complicated picture
of biasing as it could be non-linear scale dependent and stochastic
(e.g. 
Matarrese et al. 1997; 
Dekel \& Lahav 1999; 
Blanton et al. 2000; Somerville et al. 2001).

Given two unknown functions, $w(z)$ and $b(z)$, we see that clustering of galaxies with redshift
will not provide a clear test of Cosmology, unless $b(z)$ is specified 
a priori by other arguments (from galaxy formation scenarios).

\subsection{The Present Epoch Growth Factor}

At a given epoch (e.g. the present) the growth factor 
is nearly `blind' to dark energy.
In the case of Einstein's Cosmological Constant $(w=-1)$
the linear theory relation is approximately (Lahav et al. 1991):
\begin{equation}
f \equiv { {d \ln \delta} \over {d \ln a } } \approx
 \Omega_{m0}^{0.6} + { 1 \over {70}} \Omega_{Q0} (1 + \Omega_{m0} /2)
\label{eq:f_lambda}.
\end{equation}
For an epoch-independent  $w$ the growth factor (for a flat Universe) is approximately 
(Wang \& Steinhardt 1998): 
\begin{equation}
f \approx \Omega_{m0}^{\alpha(w)}
\label{eq:f_w}.
\end{equation}
with 
\begin{equation}
\alpha (w)  \approx  \frac{3}{5-w/(1-w)} 
\label{eq:f_alpha}.
\end{equation}
For a $\Lambda$CDM model ($w=-1$) it agrees reasonably well with eq. 
Eq.~(\ref{eq:f_lambda}).
The main implication of the weak dependence of $f$ on $w$ 
is that peculiar velocities 
and redshift distortion cannot constrain Dark Energy at all.
The good news is that they can tell us about $\Omega_{m0}$ independently 
of any assumptions about the nature of the Dark Energy.

\subsection{Virialization}

A well known result for the spherical collapse in an Einstein- de Sitter Universe
is that the radius $R_{vir}$ at virialization is half 
the turn-around radius $R_{ta}$.
In the presence of a Cosmological Constant $\Omega_{Q0}$ (assuming $w=-1$) 
the relation is more complicated (Lahav et al. 1991): 
\begin{equation}
{R_{vir} \over R_{ta} } \approx { {1-\eta/2} \over {2 - \eta/2} }, 
\end{equation}
where $\eta \equiv { \Omega_{Q0} \over {4 \pi G \rho_{ta} }} < 1 $
(the condition for turn around). 
The minimal possible ratio  $\frac{R_{vir}}{R_{ta}}$ is about 
$\frac{1}{3}$ (compared with $\frac{1}{2}$ in the Einstein de Sitter case), 
so it is difficult to detect the effect observationally.
Other details for spherical collapse in a Universe with 
$\Omega_{Q0} > 0$ are given in Lilje (1992) and Lokas \& Hoffman (2002),
and a  more  general case of a spherical collapse
in a Quintessence model is derived in  Wang \& Steinhardt (1998).

%
%

\section{Dark Energy constraints 
from the 2dFGRS Power spectrum combined with the  CMB}

For conventional Dark Energy models
(where the Dark Energy density was negligible at high redshift)
the matter power spectrum of fluctuations is 
actually insensitive to Dark energy.
On the other hand, the matter power spectrum derived e.g. from galaxy redshifts surveys
can constrain other parameters such as the CDM shape parameter $\Omega_m h$.
For example, the 2dF galaxy power spectrum based on 160K redshifts 
fits well a $\Lambda$-CDM model (Percival et al. 2001) 
with $\Omega_m h = 0.20 \pm 0.03$
and $\Omega_b/\Omega_m = 0.15 \pm 0.07$ (1-sigma errors)
over the scales $ 0.02 < k < 0.15 h$ Mpc$^{-1}$ 
(assuming a reasonable prior on the Hubble constant).
Some deviations from this standard model are possible, e.g. 
a contribution of massive neutrinos with an upper limit of
$\Omega_{\nu}/\Omega_m < 0.13$ (95\%
CL) is still allowed by the current 2dFGRS data  (Elgaroy et al. 2002).

To get interesting limits on Dark Energy the 2dFGRS power spectrum can
be combined with the CMB.  
Efstathiou et al. (2002) showed that 2dFGRS+CMB provide  
evidence for a positive Cosmological Constant $\Omega_{Q0}\sim 0.7$
(assuming $w=-1$), 
independently of the studies of Supernovae Ia.
As explained in Percival et al. (2002), the
shapes of the CMB and the 2dFGRS power spectra are insensitive to Dark
Energy. The main important effect of the Dark Energy is to alter
the angular diameter distance to the last scattering, and
thus the position of the first acoustic peak.  For a flat
model, the present day horizon size is given for a constant $w$ by:

\begin{equation}
r_H(z=0) = \frac{2c}{H_0} \Omega_{m0}^{\gamma(w)}
\label{eq:r_H}
\end{equation}
with 
\begin{equation}
\gamma(w)  \approx  \frac{2 w}{1-3.8 w} 
\label{eq:gamma_w}.
\end{equation}
It turns out that the CMB+2dFGRS constrain the combination of $w$ and the
Hubble constant $h$, as shown in Figure 1 for the case of a flat Universe.
With the extra HST
constraint of $h = 0.72 \pm 0.08$ (Freedman et al. 2001) and a prior
$w \geq -1$ 
Percival et al. (2002) find an upper limit 
$$
w <-0.52 \;\; (95\% CL).
$$

Lewis \& Bridle (2002) combined the CMB, 2dFGRS, HST, BBN and SN Ia and
found after marginalizing over 8 other free parameters $w < -0.68$
(95\% CL) for a flat Universe with a prior of $w \geq -1$, and $-1.6 < w <
-0.73$ (95\% CL) with no prior on $w$.  Similar results for $w$ are given
by Melchiorri et al. (2002) and others.  The main conclusion so far is that
$w=-1$ (Einstein's Cosmological Constant) is consistent with all
currently available data, but a slight deviation from it cannot yet
be ruled out, including the possibility of $w < -1$.

\begin{figure}
\centerline{\vbox{
\psfig{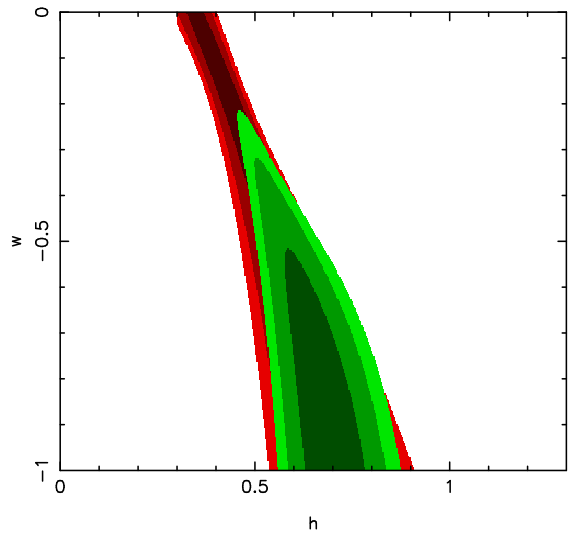}
}}
\caption[]{Likelihood contours for the (constant) Equation 
of State parameter $w$ (with prior $w\geq -1$) 
vs. the Hubble parameter $h$.
The background set of contours 
are from using 2dFGRS+CMB and the foreground contours
(at the levels of 1,2 and 3 sigma)
are based on 2dFGRS+CMB+HST.
From Percival et al. (2002).}
\end{figure}

\section{The Alcock-Paczynski Curvature Test} 

Consider a spherical object at high redshift.  If the wrong cosmology
is assumed in interpreting the distance-redshift relation along the
line of sight and in the transverse  direction, the sphere will appear
distorted.  Alcock \& Paczynski (1979) pointed out that this curvature
effect could be used to estimate the cosmological constant.  Phillips
(1994) suggested that if the correlation function in real space is
isotropic it could be used as a spherical object to measure the
effect.  Certain studies were less optimistic than others about the
possibility of measuring this A-P effect.  For example, Ballinger,
Peacock and Heavens (1996) argued that the geometrical distortion
could be confused with the dynamical redshift distortions caused by
peculiar velocities and characterized by the linear theory parameter
$\beta \equiv \frac{\Omega_m^{0.6}}{b}$.  Their model for the
distorted power spectrum $P^S(k_{\parallel}, {\bf k}_{\perp})$ as a
function of line of sight and perpendicular direction was applied by
Outram et al. (2001) to a sample of 10,000 quasars from the 2dF quasar
redshift survey.  They found best fit values with relatively large
error bars, $\beta = 0.39^{+0.18}_{-0.17}$ and $\Omega_m = 1 -
\Omega_{Q0} = 0.23^{+0.44}_{-0.13}$ (assuming $w=-1$), 
and illustrated that an
Einstein-de Sitter Universe can only marginally be rejected.
See also Popowski et al. (1998).

Matsubara \& Szalay (2002a,b) showed that the typical SDSS and 2dF
samples of normal galaxies at low redshift ($\sim 0.1$) have
sufficiently low signal-to-noise, but they are too shallow to detect
the A-P effect.  On the other hand, the quasar SDSS and 2dFGRS surveys
are at a useful redshift, but they are too sparse.  A more promising
sample is the SDSS Luminous Red Galaxies survey (out to redshift
$z\sim 0.5$) which turns out to be optimal in terms of both depth and
density.  Figure 2 shows the expected lowest error bounds (via
Fisher matrix analysis) on the Dark Energy parameters for an Equation
of State of the form $w(z) = w_0 + w_1 z$.  This analysis is based on a
 galaxy counts in 15 $h^{-1}$Mpc
spheres.  While this analysis is promising, it remains to be tested
if non-linear clustering and complicated biasing (which is quite
plausible for red galaxies) would not `contaminate' the measurement of
the Equation of State.
Even if the A-P test turns out to be less accurate than other
cosmological tests (e.g. CMB and SN Ia) the effect itself is an
important ingredient in analyzing the clustering pattern of
galaxies at high redshifts.

A variant of the A-P method has been proposed by Hui et al. (1999) and
McDonald et al. (1999) using observations of the Lyman-$\alpha$ forest
in the spectra of close quasar pairs.  The idea is to compare the
auto-correlation along the line of sight with the cross-correlation
between two (or more) close lines of sight to quasars.  It has been
estimated that 30 or so pairs of quasar spectra are needed to
estimate of the Dark Energy parameters with reasonable accuracy,
subject to modelling uncertainties.

\begin{figure}
\centerline{\vbox{
\psfig{figure=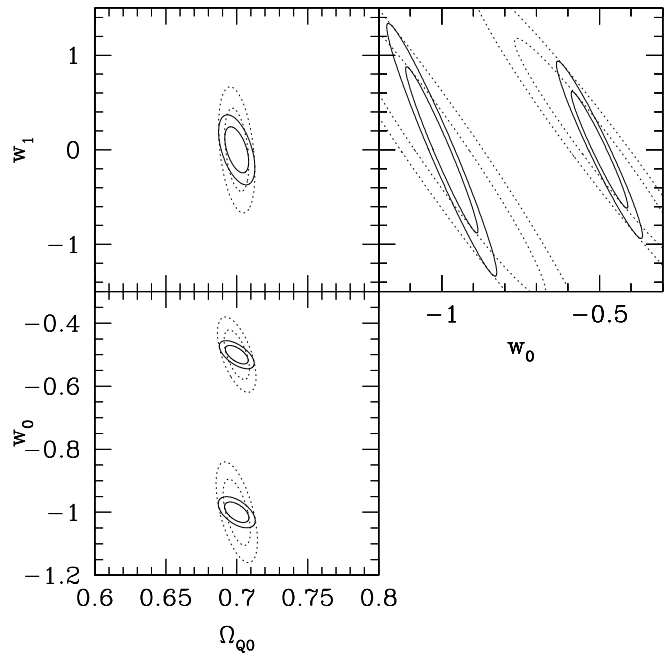,height=9.cm}
}}
\caption[]{ Expected (minimal) error bounds on the dark energy
parameters ($w_0, w_1, \Omega_{Q0}$) from the Luminous Red Galaxies
SDSS survey. One of three parameters are held fixed.  Inner ellipses
represent the 1-sigma uncertainty level of one-parameter distribution
function. Outer ellipses represent the 1-sigma of the joint
probability distribution.  The biasing parameter is held fixed (solid
lines) or is marginalized over (dotted lines).  From Matsubara \&
Szalay (2002b).}
\end{figure}

\section{Abundance of Clusters and Galaxies with Redshift} 

A popular method for constraining cosmological parameters is to count
the number $N(z)$ of clusters with redshift.  Commonly the
Press-Schechter mass function (or one of its variants) is used to
predict the number of clusters with redshift.  The Equation of State
parameter is mainly sensitive to the growth rate $\delta(z)$ and the
volume element (e.g. Wang \& Steinhardt 1998; Weller et al. 2002).
The main uncertainty is in relating the observed cluster quantities
such as temperature, velocity dispersion or Sunyaev-Zeldovich effect to
the cluster mass. While some empirical relations exist, they suffer
from various systematic effects, e.g. evolution.  The cluster
abundance test is discussed in more detail by others in this Volume.

Another variant of this method has been suggested by Newman \& Davis (2002)
for the abundance of galaxies
(e.g. in  the DEEP2 survey, over the redshift range $0.7 < z < 1.5$ )
as a function of their circular velocity.
Subject to possible systematic errors, 
this test can provide information about
Dark Energy parameters, in particularly if extra constraints
on $\Omega_m$ are available.

\section{Cross Correlation of the CMB and Galaxy Surveys}

An entirely different method to constrain Dark Energy, by cross correlating 
the CMB with tracers of the mass density at low redshift,
has been proposed by Crittenden \& Turok (1996).
The idea is that in a Universe with a Cosmological Constant 
the gravitational potential is time dependent even on large scales
(unlike in the 
Einstein de Sitter case, in which the potential is constant with time). 
Then CMB fluctuations can arise via the Integrated Sachs Wolfe (ISW) effect,
as the photons travel through the time-dependent potentials.
They can be observed by searching for spatial correlations between 
the CMB and the local matter density (e.g. from galaxy counts $N$),
$< \delta N \delta T >$.
Boughn \& Crittenden (2002) crossed-correlated the CMB with distant radio 
sources, and could only place an upper limit on the cross-correlation 
signal. For a Cosmological Constant model 
($w=-1$) they translated it to an upper limit of 
$$
\Omega_{Q0} \leq 0.74 \;\; (95 \% CL) \;.
$$
This is actually the only measurement discussed in this review which 
argues marginally against a Dark Energy component
(it is actually similar to the upper limit derived from the
frequency of lensed quasars; Kochaneck 1996).
However, this upper limit is still in accord with the concordance
model, and systematic effects such as galaxy biasing may confuse the 
deduction of the cosmological parameters.
If indeed a Dark Energy component does exist, it should be detectable
with future CMB maps and galaxy surveys.

\section{Conclusions}

The main conclusions of this review are:

\begin{itemize}
 
\item  The present epoch matter power spectrum and dynamics
are almost `blind' to a Dark Energy component.
This is actually useful as these measurements can be used
to deduce the matter density parameter $\Omega_m$ independent 
of any assumption  on the nature of the Dark Energy, 
and then combined with other probes such as the CMB and Supernovae Ia.

\item Geometrical tests and counts at high redshift 
are potentially useful, but they suffer 
from degeneracy with other parameters such as $\Omega_m$ and biasing.

\item The current data are consistent 
with $w=-1$ (i.e. a Cosmological Constant), but other forms 
of Quintessence $w(z)$ are still possible.

\end{itemize} 

\section{Outlook}

There is general acceptance (perhaps too strongly) of the
`concordance' model with the following ingredients: 4\% baryons, 26\%
Cold Dark Matter (possibly with a small contribution of massive
neutrinos) and the remaining 70\% in the form of Dark Energy (the
Cosmological Constant or `Quintessence').  Conceptually, it seems we
have to learn to live in a multi-component complex Universe, which
perhaps takes us away from an idealized model motivated by Occam's
razor.

While phenomenologically the $\Lambda$-CDM
model has been successful in fitting a wide range
of cosmological data, there are some open questions:

\begin{itemize}

\item
Both components of the model, $\Lambda$ and CDM, 
have not been directly measured.
Are they `real' entities or just `epicycles'?
Historically  epicycles were actually quite useful 
in forcing observers to improve their measurements and theoreticians to think
about better models!

\item
`The Old Cosmological Constant problem': 
Why is $\Omega_{Q0}$ at present so small relative to what is expected
from Early Universe physics? 

\item 
`The New Cosmological Constant problem': 
Why is $\Omega_{m0} \sim \Omega_{Q0}$ at the present-epoch?
Why is $w \approx -1$ ?
Do we need to introduce new physics 
or to invoke the Anthropic Principle to explain it?

\item
There are still open problems in 
$\Lambda$-CDM on the small scales,
e.g. galaxy profiles and satellites.

\item 
The age of the Universe is uncomfortably close to some estimates for the 
age of the Globular Clusters, if their epoch of 
formation was late. 

\item  
Could other (yet unknown) models fit the data equally well?

\item
Where does the field go from here?
Should the activity focus on refinement of 
the cosmological parameters within $\Lambda$-CDM, 
or on introducing entirely new paradigms?

\end{itemize} 


%
%

\acknowledgements{I am grateful to members of the 2dF galaxy redshift
survey team and participants of the Leverhulme Quantitative Cosmology group
in Cambridge for helpful discussions.}

\vfill
\end{document}